\documentclass[printer]{aa}

\bibpunct{(}{)}{;}{a}{}{,}

\usepackage{graphicx}
\usepackage{txfonts}
\usepackage{soul}
\usepackage{txfonts}
\usepackage{color}
\usepackage{ulem}

\newcommand{\hii}{H\textsc{ii}}
\def\ks{km s$^{-1}$}

\def\s{$^{\prime\prime}$}

\def\cm3{cm$^{-3}$}

\def\2{$^{12}$CO}
\def\3{$^{13}$CO}
\def\8{C$^{18}$O}
\def\msol{M$_\odot$}

\def\cm2{cm$^{-2}$}

\begin{document}

\title{Unveiling the substructure of the massive clump AGAL G035.1330$-$00.7450}

\author {M. E. Ortega \inst{1}
\and A. Marinelli \inst{1}
\and N. L. Isequilla \inst{1}
\and S. Paron \inst{1}
}

\institute{CONICET - Universidad de Buenos Aires. Instituto de Astronom\'{\i}a y F\'{\i}sica del Espacio
             CC 67, Suc. 28, 1428 Buenos Aires, Argentina\\
             \email{mortega@iafe.uba.ar}
}

\offprints{M. E. Ortega}

   \date{Received <date>; Accepted <date>}

\abstract{It is known that high-mass stars form as result of the fragmentation of massive molecular clumps. However, what is not clear is whether this fragmentation gives rise to cores massive enough to form directly high-mass stars or leads to cores of low and intermediate masses that generate high-mass stars, acquiring material from their environment.}
{Detailed studies towards massive clumps at the early stage of star formation are needed to collect observational evidence that shed light on the fragmentation processes from clump to core scales. The infrared-quiet massive clump AGAL G035.1330$-$00.7450 (AGAL35) located at a distance of 2.1~kpc, is a promising object to study both the fragmentation and the star formation activity at early stages.}
{Using millimeter observations of continuum and molecular lines obtained from the Atacama Large Millimeter Array database at Bands 6 and 7, we study the substructure of the source AGAL35. The angular resolution of the data at Band 7 is about 0\farcs7, which allow us to resolve structures of about 0.007 pc ($\sim$1500 au).} 
{The continuum emission at Bands 6 and 7, shows that AGAL35 harbours four dust cores, labeled from C1 to C4, with masses below 3~\msol. The cores C3 and C4 exhibit well collimated, young, and low-mass molecular outflows related to molecular hydrogen emission-line objects, previously detected. The cores C1 and C2 present CH$_3$CN J=13--12 emission, from which we derive rotational temperatures of about 180 and 100~K, respectively. These temperatures allow us to estimate masses of about 1.4 and 0.9~\msol~for C1 and C2, respectively, which are about an order of magnitude smaller than those estimated in previous works and are in agreement with the Jeans mass of this clump. In particular, the moment 1 map of CH$_3$CN emission suggests the presence of a rotating disk towards C1, which is confirmed by the CH$_3$OH and CH$_3$OCHO (20--19) emissions. On the other hand, the CN N=2--1 emission show a clumpy and filamentary structure that seems to connect all the cores. These filaments might be tracing the remnant gas of the fragmentation processes taking place within the massive clump AGAL35  or the gas that is being transported towards the cores, which would imply a competitive accretion scenario.} 
{The massive clump AGAL35 harbours four low-intermediate mass cores with masses below 3~\msol, about an order of magnitude smaller than the estimated in previous works. This study shows that in addition to the importance of high-resolution and sensitivity observations for a complete detection of all fragments, it is very important to accurately determine the temperature of such cores for a correct mass estimation. Finally, although no high-mass cores were detected towards AGAL35, the presence of a filamentary structure connecting all the cores, leaves open the possibility of high-mass stars forming through the competitive accretion mechanism. }

\titlerunning{.}
\authorrunning{M. E. Ortega et al.}


\maketitle
%

\section{Introduction}

It is known that massive stars ($\geq$\,8~\msol) form  in  clusters  deeply  embedded in massive molecular clumps.
Nowadays there are evidence pointing to that such clumps, or some of them, can be density-amplified hubs of converging molecular filaments systems, in which according to \citet{kumar20}, low-mass stars 
form along the filaments, while the high-mass stars do in the hubs. These clumps/hubs may collapse under self-gravity and fragment into multiple cores. The number of cores and their mass distribution depend on the processes that regulate the fragmentation, which are still a matter of vigorous discussion and debate \citep{moscadelli2021,palau2018}.  The mass distribution of the fragments together with the mass reservoir available for the formation of individual stars could 
give us information to determine whether the massive-star formation is due to an individual monolithic core collapse, or to a global hierachical collapse of a clump \citep{motte18}. As it was proposed some years ago,  the first case, the `core-accretion' model \citep{mckee02}, predicts the existence of quasi-equilibrium massive cores providing all the material to form high-mass stars, and in the second one, the `competitive accretion' model (e.g. \citealt{bonnell08}), the parent molecular clump fragments into many low-mass cores that competitively accrete the surrounding gas. In any case, nowadays it is clear that the high-mass star-formation scenarios are ruled by
processes that are not quasi-static but simultaneously evolve with both cloud and cluster formation \citep{motte18}. 

The fragmentation of a molecular cloud/filament that leads to different mass distributions of fragments
depends on the evolutionary stage of the cloud and the spatial scale considered. As \citet{kai13}
point out, an explanation for the different fragmentation characteristics can be the size-scale-dependent collapse time-scale that results from the finite size of real molecular clouds, which is indeed predicted by analytical models \citep{pon11}. 

Regarding to the mass distribution of such molecular fragments, recent studies with spatial resolutions of 0.005--0.02 pc, find a large number of low-mass ($\lesssim$ 1 \msol) molecular fragments in infrared-quiet massive clumps located in high-mass star-forming regions (\citealt{palau21,li21,sanhueza2019}). On the other hand, \citet{neupane2020} and \citet{csengeri2017}, with similar spatial resolutions, reported limited fragmentation (very few cores and with superJeans masses,
well above the solar mass)  
in regions at the early stages of high-mass star formation. Therefore, it is not clear if the observation of limited fragmentation has a physical origin or it is due to an observational issue. For instance, \citet{hen17}, based on high resolution and sensitivity
observations, studied a  massive clump of about 100~\msol~located within the infrared dark cloud (IRDC) G035.39$-$00.33, and found several low-intermediate mass cores within 2--8 \msol~mass range. They concluded that the previously reported dearth of such low-mass objects in this massive clump was due to an observational, rather than physical origin, and that some low-mass cores form coevally in the neighbourhood of more massive objects. Thus, it is essential to carry out more studies about fragmentation of massive clumps in order to shed light on the high-mass star formation models.



 \begin{figure*}[h]
   \centering
   \includegraphics[width=18cm]{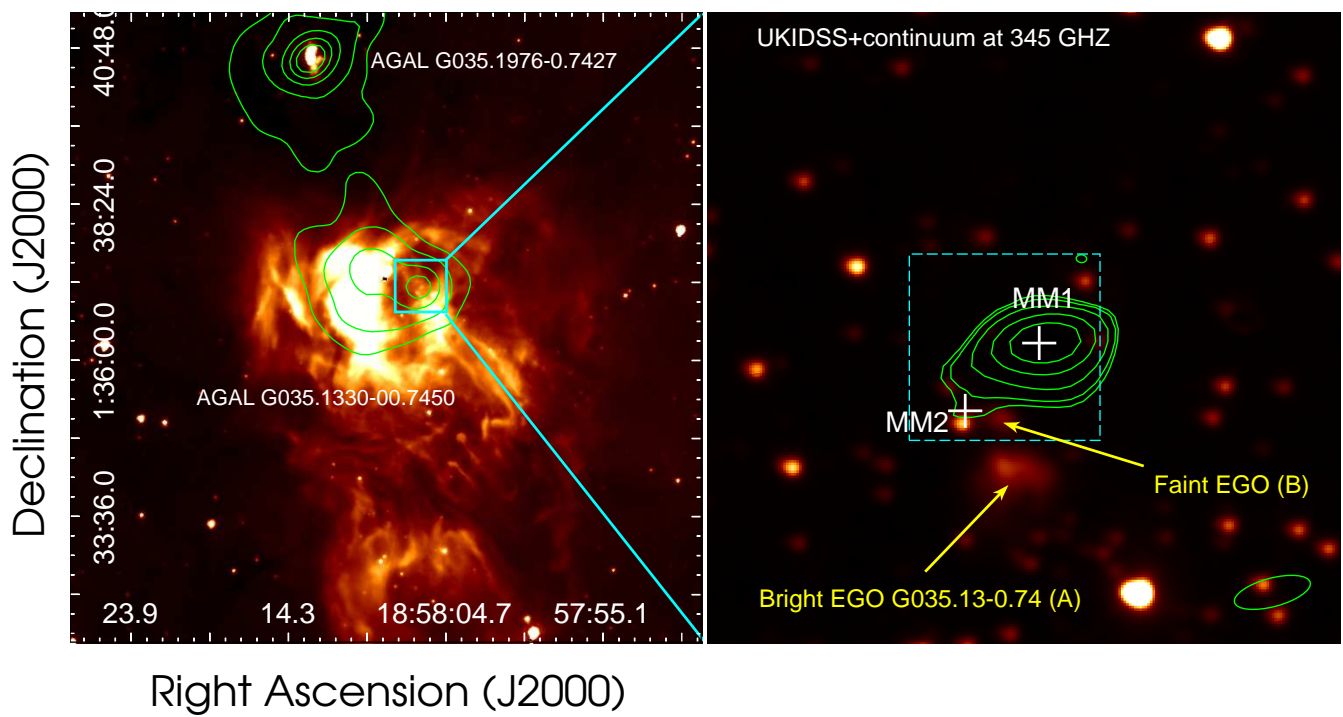}
    \caption{Left. Overview of the \hii~region G035.126$-$00.755 at {\it Spitzer}-IRAC 8.0~$\mu$m image. The green contours represent the ATLASGAL emission at 870~$\mu$m. Levels are at 0.7, 1.2, 2.6, 4.5, and 7~Jy beam$^{-1}$. Right. Close--up view of the massive clump AGAL35 at UKIDSS K-band image. The green contours represent the ALMA continuum emission at 345~GHz (7~m array from \citet{csengeri2017}). Levels are at 0.05, 0.10, 0.20, 0.35, and 0.50 Jy beam$^{-1}$. The beam is indicated at the bottom right corner. The white crosses indicate the position of the two fragments found by \citet{csengeri2017} towards AGAL35. The dashed line square indicates the region studied in this work.}
    \label{intro}
\end{figure*}

To better understand how the fragmentation and accretion processes at clump scale occur, it is crucial to study massive molecular clumps at the earliest stages of star formation with  high angular resolution and sensitivity.  Thus, we decided to look for some IR-quiet source with signs of star-forming activity that have interferometric observations. Analyzing the work of \citet{froebrich2011}, in which several hydrogen emission-line objects (MHOs), signs of outflows driven by massive young stellar objects, were detected near the cluster Mercer\,14, 
we realize that some of them coincide with the IR-quiet massive molecular clump AGAL~G035.1330$-$00.7450 (hereafter AGAL35).  According to \citet{csengeri2017}, this IR-quiet source presents limited fragmentation, and hence, given that there are related high-angular resolution data in the Atacama Large Millimeter Array (ALMA) database, we conclude that this is a very appropriate source to study the early massive star-forming processes. 

In this work, we present an analysis of the fragmentation and the kinematics of the molecular gas of the massive clump AGAL35 in order to unveil its internal structure. The paper is organized as follows: Sect.\,\ref{present} presents the source AGAL35, Sect.\,\ref{ALMAdata} describes the used data, in Sects.\,\ref{results} and \ref{discussion} we present results and discussion, respectively, and finally Sect.\,\ref{concl} states our concluding remarks.

\section{Presentation of the region}
\label{present}

Figure \ref{intro}, left, shows a {\it Spitzer}-IRAC image at 8~$\mu$m of the \hii~region G035.137$-$00.762. \citet{anderson2015}, based on a radio recombination line, estimated a systemic velocity of 34.5~\ks~for the \hii~region, which corresponds to a kinematic distance of about 2.1~kpc.  The source AGAL35 appears located in projection towards the bulk of the 8~$\mu$m emission. \citet{wienen2012}, based on ammonia (and $^{13}$CO) lines, found towards AGAL35 velocity components at 32 (33) and 35 (36)~\ks, with kinetic temperatures of about 20 and 25~K, respectively.  \citet{csengeri2017a} include AGAL35 in a study of an unbiased sample of infrared-quiet massive clumps in the Galaxy that potentially represent the earliest stages of massive cluster formation, and estimated a mass of about 466~\msol~for this clump.

Figure \ref{intro}, right, shows a close-up view of the central region of AGAL35 at the near-infrared K-band obtained from UKIDSS\footnote{http://wsa.roe.ac.uk/index.html}. The green contours represent the ALMA 7~m continuum emission at 345~GHz presented by \citet{csengeri2017}. The authors identified two fragments labeled MM1 and MM2, with masses of about 36 and 8~\msol~(considering a dust temperature of about 25~K), respectively. 

It can be noticed two near-infrared extended sources, labeled `Bright' (A) and `Faint EGO' (B) \citep{froebrich2011}, which are in positional coincidence with the extended green object EGO G035.13$-$0.74 (hereafter EGO35) catalogued by \citet{cyganowski2008}. Both extended sources are located towards the south of the submillimeter emission. 

Finally, \citet{froebrich2011} found several molecular hydrogen emission-line objects (MHOs) in the region (see Fig. \ref{MHOs_CO}) and, based on GLIMPSE infrared color criteria, identified possible associated central sources.

\section{Data}
\label{ALMAdata}

Data cubes were obtained from the ALMA Science Archive\footnote{http://almascience.eso.org/aq/}.
We used data from two projects: 2015.1.01312 (PI: Fuller, G.) and 2013.1.00960 (PI: Csengeri, T.), whose observation dates were 2016-03-21 and 2015-04-03, respectively.
The observed frequency ranges and the spectral resolutions are: 224.24--242.75~GHz and 1.3~MHz (Band 6), and 333.35 to 348.97~GHz  and 1.9~MHz (Band 7) for each project, respectively. 

The single pointing observations for this target were carried out using the following telescope configurations with Min/Max Baseline(m): 15/460 for project 2015.1.01312 and 15/327.8 for project 2013.1.00960, in the 12~m array in both cases. The maximum recoverable scales at Band 6 and Band 7 are 5.9 and 6.7~arcsec, respectively. 

It is important to remark that even though the data of both projects passed the QA2 quality level, which assures a reliable calibration for a “science ready” data, the automatic pipeline imaging process may give raise to a clean image with some artefacts. For example, an inappropriate setting of the parameters of the {\it clean} task in CASA could generate artificial dips in the spectra. Thus, we reprocessed the raw data using CASA 4.5.1 and 4.7.2 versions and the calibration pipelines scripts. Particular care was taken with the {\it clean} task. The images and spectra obtained from our data reprocessing are very similar to those obtained from the archival.

The task {\it imcontsub} was used to subtract the continuum from the spectral lines in project 2015.1.01312 using a first order polynomial. Table\,\ref{data} presents the main data parameters. The continuum images at 239 and 334~GHz are corrected for primary beam.

Given that high-spatial resolution is required to perform this study, it is important to remark that the beam size of the 334~GHz data cube provides a spatial resolution of about 0.007~pc ($\sim$ 1500 au) at the distance of 2.1~kpc, which is appropriate to characterize the substructure of AGAL35. 

\begin{table}
\tiny
\centering
\caption{Main data parameters.}
\label{data}
\begin{tabular}{lcccc}
\hline\hline
                  & beam size & FoV & $\Delta$v & rms  Noise          \\
                  &   (\s $\times$ \s)             & (arcsec) & (\ks) &(mJy beam$^{-1}$)\\
\hline                 
Proj. 2013.1.00960 & & & &\\
\hline
Cont. 334 GHz     &  0.80 $\times$ 0.66  & 17 & $-$ & 5       \\
$^{12}$CO J=3--2  &  0.80 $\times$ 0.66  & 17 & 0.9 & 15       \\
\hline
Proj. 2015.1.01312 &  & & &\\
\hline
Cont. 239 GHz     &  0.78 $\times$ 0.62  & 25 & $-$ & 1       \\
C$^{17}$O (2--1) & 0.83 $\times$ 0.66 & 25 & 1.5 & 7 \\
CH$_{3}$CN (13--12) & 0.78 $\times$ 0.62 & 25 & 1.4 & 8 \\
CH$_{3}$OCHO (20--19) & 0.82 $\times$ 0.64 & 25 & 1.5 & 6 \\
CH$_{3}$OH (20--19)  & 0.83 $\times$ 0.66   & 25  & 1.5  & 7 \\ 
CN (2--1) & 0.82 $\times$ 0.64 & 25 & 1.5 & 6 \\
\hline
\hline
\end{tabular}
\end{table}

\section{Results}
\label{results}

In the following subsections we present the results from the analysis of the ALMA data towards AGAL35.

\begin{table*}
\tiny
\caption{Main dust cores parameters at 239 and 334~GHz using a 2D Gaussian fitting from CASA software.}
\label{contparams}
\centering
\begin{tabular}{ccccccccccc}
\hline\hline
Core & RA & Dec. & $\Theta_{\rm maj}^{\dagger}$ & $\Theta_{\rm min}^{\dagger}$  &  Peak int.$_{\rm 239}$ & S$_{239}$ & Mass$_{239}$ (25~K) & Peak int.$_{\rm 334}$ & S$_{\rm 334}$ & Mass$_{\rm 334}$ (25~K)\\
 & (J2000) & (J2000) & (arcsec)  &  (arcsec) & mJy beam$^{-1}$ & mJy & (\msol) & mJy beam$^{-1}$ & mJy & (\msol)\\
\hline
C1 & 18:58:06.1 & +01:37:07.1 & 1.58  & 0.97 & 74 & 267 & 9.4 & 241 & 710 & 13.3\\ 
C2 & 18:58:06.3 & +01:37:07.4 & 1.31  & 0.85 & 26 & 186 & 6.5 & 89 & 410 & 7.7\\ 
C3 & 18:58:06.6 & +01:37:05.0 & 0.84  & 0.78 & 33 & 37 & 1.3 & 80 & 81 & 1.5\\ 
C4 & 18:58:06.6 & +01:37:02.2 & 0.93  & 0.88 & 31 & 59 & 2.1 & 101 & 173 & 3.1\\ 
\hline
\multicolumn{11}{l}{$^{\dagger}$ Beam deconvolved.} 
\end{tabular}
\end{table*}

\subsection{Continuum emission at 239 and 334~GHz: tracing the fragmentation of AGAL35}  
\label{dust}

\begin{figure}[h]
   \centering
   \includegraphics[width=9cm]{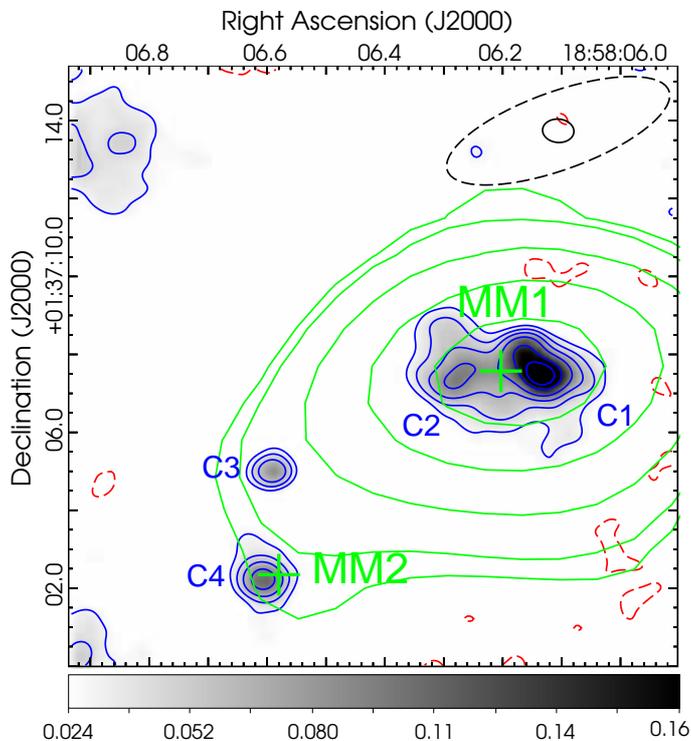}
    \caption{ALMA continuum emission at 334~GHz (12~m array). The colour scale unit is Jy beam$^{-1}$. The blue contours levels are at 15, 30, 50, 80, 140, and 200~mJy beam$^{-1}$. The green contours correspond to the ALMA continuum emission at 345~GHz (7~m array) reported by \citet{csengeri2017}. The red dashed contours correspond to $-$15~mJy beam$^{-1}$. The beams of both observations are indicated at the top right corner.}
              \label{334GHz}
    \end{figure}

Figure \ref{334GHz} shows the ALMA continuum emission at 334~GHz (12~m array) in grayscale and blue contours. The green contours represent the ALMA continuum emission at 345~GHz (7~m array) presented in the work of \citet{csengeri2017}. The authors identified two dust condensations related to AGAL35, labeled MM1 and MM2, whose positions are indicated by the green crosses. The better angular resolution of the 334~GHz observations allows us to identify four dust cores towards AGAL35, which are labeled from C1 to C4. The condensation MM1 appears fragmented into cores C1 and C2, while the condensation MM2 seems to be associated with core C4. It also can be noticed two incompletely mapped structures towards the northeast and southeast border of the continuum map. 

\begin{figure}[h]
   \centering
   \includegraphics[width=9cm]{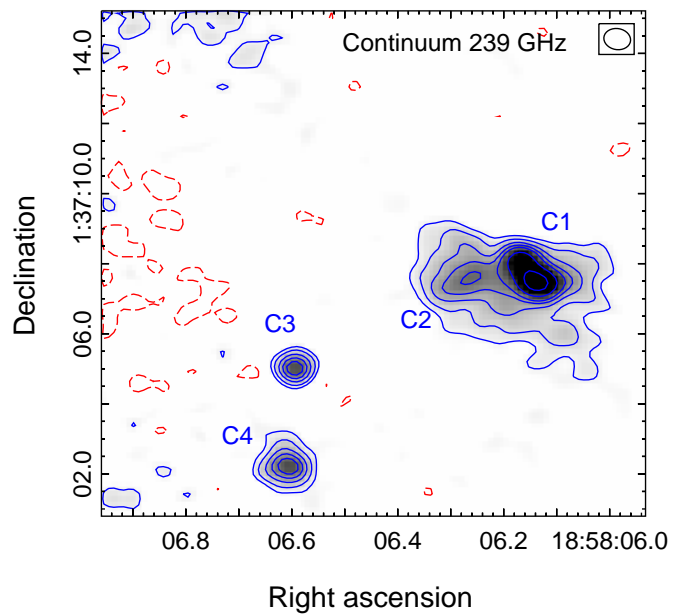}
    \caption{Continuum emission at 239~GHz extracted from the CH$_3$CN J=13--12 datacube. The grayscale goes from 3 to 45~mJy beam$^{-1}$. The blue contours levels are at 3, 8, 14, 20, 26, 40, and 65~mJy beam$^{-1}$. The red dashed contours correspond to $-$3~mJy beam$^{-1}$. The beam is indicated at the top right corner.}
              \label{239GHz}
    \end{figure}

Figure \ref{239GHz} shows the continuum emission at 239~GHz extracted from the CH$_3$CN J=13--12 datacube, where  the same four molecular cores can be appreciated. It is important to mention that at 239~GHz, the contamination of the free-free continuum emission might be not negligible. For example, \citet{isequilla21} found that towards the core MM1 in the IRDC G34.43$+$00.24, the contribution of the free-free continuum emission is about 15 percent at 93~GHz and negligible at 334~GHz. In AGAL35, while the absence of centimeter radio continuum data prevent us from estimate the free-free emission contribution, the absence of UKIDSS sources associated with the cores (see Section \ref{mho_disc}), which indicates the youth of the sources, suggests that the contribution of the ionized gas emission should not be important towards this region.

Table \ref{contparams} presents the main parameters of the dust continuum cores observed at 239 and 334~GHz. Columns 2 and 3 give the absolute position, Cols.\,4 and 5 present the major and minor axis related to the source size, respectively, from a 2D Gaussian fitting, Cols.\,6, 7, and 8 show the peak intensity, the integrated intensity, and the mass, respectively, at 239~GHz, Cols.\,9, 10, and 11 show the same parameters at 334~GHz. Despite, the cores sizes are at least a factor four lower than the maximum recoverable scales at both frequencies, we can not discard an underestimation in the calculation of the masses due to the missing flux of the interferometric observations.   

The mass of gas for each core was estimated from the dust continuum emission at 239 ($\lambda \sim$ 1.3~mm) and at 334~GHz ($\lambda \sim$ 0.9~mm) following  \citet{kau08},

\begin{eqnarray}
M_{gas}=0.12~{\rm M_\odot} \left[exp\left(\frac{1.439} {(\lambda/{\rm mm})(T_{dust}/10~{\rm K})}\right)-1\right] \\ \nonumber \times\left(\frac{\kappa_{\nu}}{0.01~{\rm cm}^2~{\rm g}^{-1}}\right)^{-1}\left(\frac{S_{\nu}}{\rm Jy}\right)\left(\frac{d}{100~{\rm pc}}\right)^2\left(\frac{\lambda}{\rm mm}\right)^3
\label{massdust}
\end{eqnarray}

\noindent where $T_{dust}$ is the dust temperature and $\kappa_{\nu}$ is the dust opacity per gram of matter at 870~$\mu$m, for which we adopt the value of 0.0185~cm$^2$g$^{-1}$ \citep[][and references therein]{csengeri2017a}.

Assuming thermal coupling between dust and gas ($T_{dust}$=$T_{kin}$), and considering a $T_{kin}$ of about 25~K estimated by \citet{wienen2012} from the ammonia emission towards AGAL35, we derive masses that goes from 1.3 to 9.4~\msol~at 239~GHz, and from 1.5 to 13.3~\msol~at 334~GHz (see Table \ref{contparams}).

\subsection{$^{12}$CO J=3--2 transition: molecular outflow activity}
\label{12CO}

\begin{figure}
   \centering
   \includegraphics[width=9cm]{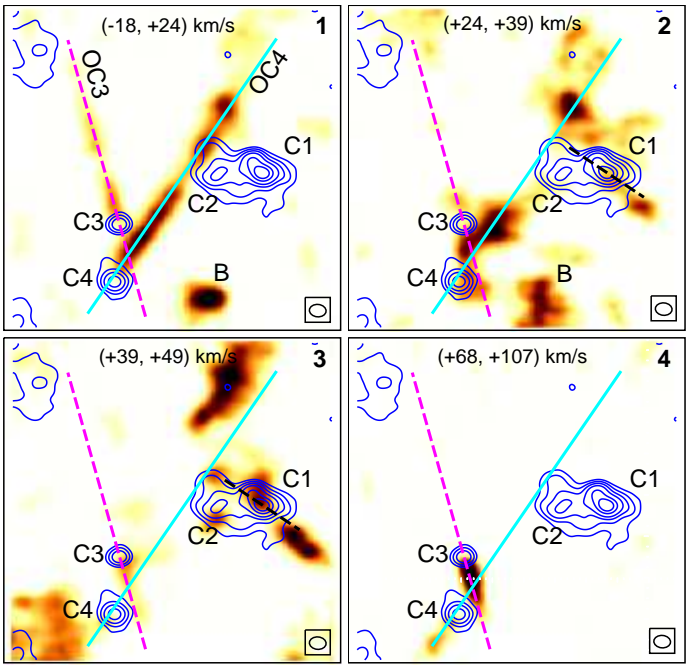}
    \caption{Integrated velocity channel maps of the $^{12}$CO J = 3$-$2 emission. The integration velocity range is exhibited at the top of each panel. The systemic velocity of the gas related to the clump AGAL35 is about 34.5~\ks. Colour-scale goes from 0.8 to 16~Jy beam$^{-1}$~\ks. Blue contours represent the continuum emission at 334~GHz. Levels are at 15, 30, 50, 80, 140, and 200~mJy beam$^{-1}$. The magenta, and cyan lines indicate the direction of molecular outflows candidates related to the cores C3 and C4, respectively. The black dashed line, in panels 2 and 3, indicates the direction of the possible rotating disk related to the core C1. The beam is indicated at the bottom right corner of each map. }
             \label{coint}
    \end{figure}

Figure \ref{coint} shows the integrated velocity channel maps of the $^{12}$CO J=3$-$2 emission with the continuum emission at 334~GHz superimposed in blue contours. The integration velocity ranges are shown at the top of each panel. The systemic velocity of the complex is about $+$34.5~\ks. 

Panel 1 shows two straight $^{12}$CO filaments remarked with magenta and cyan lines, that are clearly connected to cores C3 and C4, respectively. Given  the morphology, the location, and the velocity interval in which these molecular filaments extend, they might correspond to the blue lobes of molecular outflows arising from each core. The structure labeled B  corresponds to molecular gas likely related to EGO35, that lies outside the field of view. In panels 2 and 3 the black dashed line shows the direction of a possible rotating disk related to the core C1 (see Sections \ref{CH3CN} and \ref{molec_results}). In both panels, it can be noticed molecular gas associated with C1, and particularly in panel 3, it can be appreciated a short filament which seems to be arising from C1 and could be indicating molecular outflow activity. Given that the direction of such a structure is almost the same as the direction of the possible rotating disk (see Section \ref{CH3CN}), the hypothesis that it is a molecular outflow weakens. However, although the CH$_{3}$CN kinematics towards C1 seems to be consistent with a rotating disk, the direction of the outflow candidate might suggest the presence of other embedded source in C1, with which the interpretation of the CH$_{3}$CN spectrum as multiple velocity components cannot be discarded. Indeed further studies are needed to clarify this. Additionally, in panel 2, a curved filament that seems to connect cores C2, C3, and C4 can be appreciated. Panel 3 shows three molecular features located in projection on the condensation MM1.  Finally, panel 4 shows partially, two filament-like structures connected to cores C3 and C4, which may be associated with the red lobes of the respective molecular outflows. 

\begin{figure*}[h]
   \centering
    \includegraphics[width=18cm]{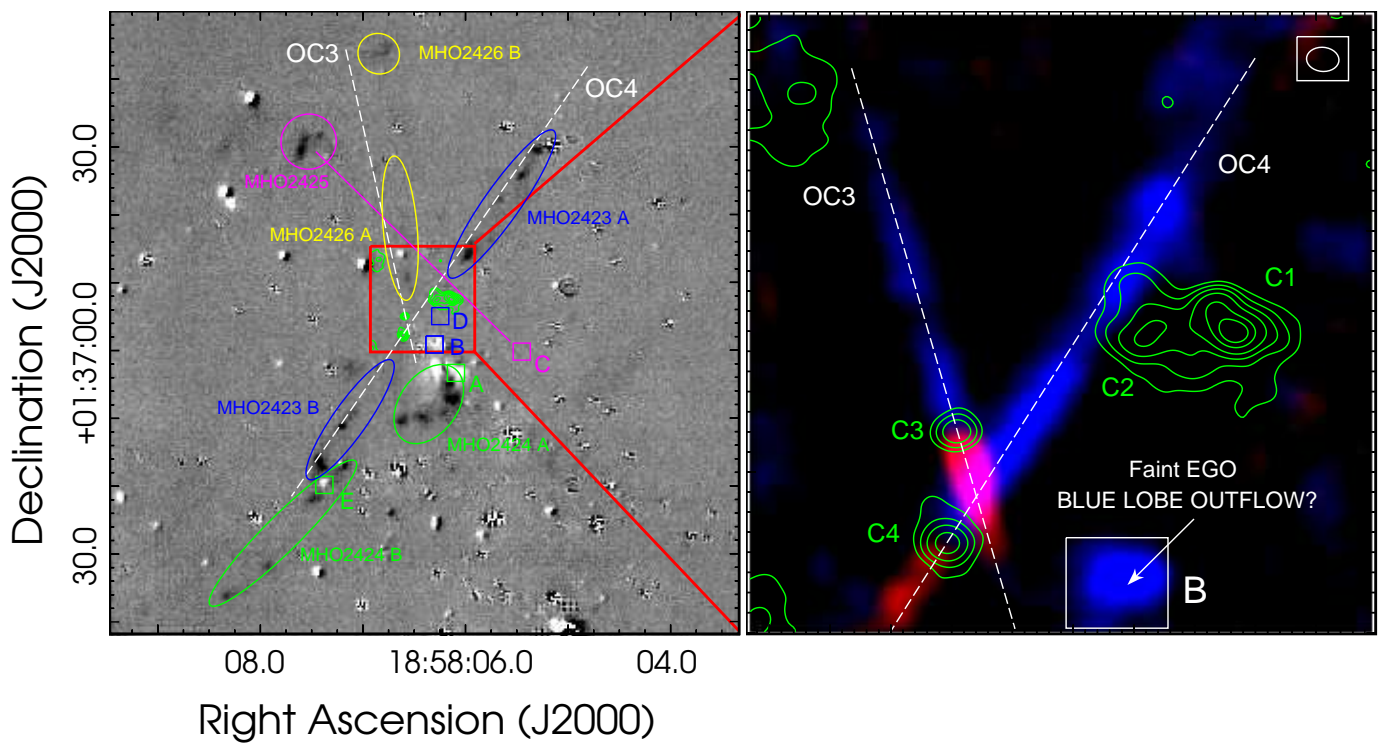}
    \caption{Left. Free continuum molecular hydrogen 1--0 S(1) emission line at 2.12~$\mu$m extracted from the  UWISH2 survey. Green contours represent the ALMA continuum emission at 334~GHz. The red square shows the field of view of the observations presented in this work. The green, blue, and magenta squares, labeled from A to E, indicate the position of the powering source candidates of the MHOs, following \citet{froebrich2011}. Right. $^{12}$CO J=3--2 emission distribution integrated between $-$18 and +24~\ks (blue) and between +68 and +107~\ks (red). Green contours represent the ALMA continuum emission at 334~GHz. The beam is indicated at the top right corner.}
              \label{MHOs_CO}
\end{figure*}

Figure \ref{MHOs_CO}, right, displays the $^{12}$CO J=3--2 emission distribution integrated between $-$18 and $+$24~\ks~(in blue), and between $+68$ and $+107$~\ks~(in red). It can be noticed the molecular outflows associated with cores C3 and C4, labeled OC3 and OC4, respectively. The respective red lobes appear to be incomplete. The `blue' condensation to the south (labeled B) might correspond to the blue lobe of the molecular outflow related to EGO35.

In order to roughly estimate the masses of the blue lobes of the  molecular outflows OC3 and OC4 (the red ones are probably incomplete), we calculate the H$_2$ column density following \citet{bertsch93}:

\begin{equation}
{\rm N(H_2)=2.0 \times 10^{20} \frac{{\rm W(^{12}CO)}}{K~kms^{-1}} (cm^{-2})},
\label{nh2}
\end{equation}

\noindent where W($^{12}$CO) is the $^{12}$CO J=3--2 integrated intensity at the the corresponding velocity intervals. The W($^{12}$CO) units were converted using:

\begin{equation}
{\rm T[K]=1.22 \times 10^3  \frac{I[mJy/beam]}{\nu^2[GHz] ~\theta_{maj}[arcsec]~\theta_{min}[arcsec]}}.   
\label{mJytoK}
\end{equation}

\noindent where $\theta_{maj}$ and $\theta_{min}$ correspond to the major and minor axis of the beam, respectively. Then, the mass of each outflow is derived from:

\begin{equation}
{\rm M=\mu~m_H~D^2~d\Omega\sum_{i}~N_i(H_2)},
\label{mass}
\end{equation}

\noindent where d$\Omega$ is the solid angle subtended by the beam size, ${\rm m_H}$ is the hydrogen mass, ${\rm \mu}$ is the mean molecular weight, assumed to be 2.8 by taking into account a relative helium abundance of 25 \%, and D is the distance. N${\rm _i}$(H${\rm _2}$) is the molecular hydrogen column density (from eq. \ref{nh2}) obtained from the $^{12}$CO J=3--2 integrated intensity map (from $-$18 to +24~\ks), considering a beam area taken over each blue lobe structure. The summation result from covering the whole extension of each blue lobe (see Fig. \ref{MHOs_CO}) with the respective beam area.Table \ref{outflow} shows the length, the mass, the momentum (${\rm P = M \bar v}$), and the dynamical age (${\rm t_{dyn} = Length/v_{max}}$) of each lobe, where ${\rm \bar v}$ and ${\rm v_{max}}$ are the median and maximum velocities of each interval velocity with respect to the systemic velocity of the gas associated with G35 complex, respectively. 

\begin{table}
\centering
\caption{Main parameters of the molecular outflows related to the cores C3 and C4.}
\label{outflow}
\begin{tabular}{lcc}
\hline\hline
                  & C3 blue lobe & C4 blue lobe \\
\hline                
              
Length (pc)       & 0.08 & 0.11   \\
Mass (${\rm M_{\odot}}$)  & 0.2 & 0.7  \\
Momentum (${\rm M_{\odot}~km~s^{-1}}$) & 6.2 & 24.8 \\
Dynamical age (10$^3$ yrs) & 1.5 & 2.0  \\
\hline

\end{tabular}
\end{table}

Figure \ref{MHOs_CO}, left, shows the free continuum molecular hydrogen 1--0 S(1) emission line at 2.12~$\mu$m extracted from the  UWISH2 survey \citep{uwish2011}. It can be noticed the perfect alignment among MHO2423 A and B, and the molecular outflow OC4, and among MHO2426 A and B, and the molecular outflow OC3.

\subsection{Analysis of the CH$_{3}$CN emission: characterizing the cores C1 and C2}
\label{CH3CN}

CH$_3$CN J=13--12 emission is observed towards the cores C1 and C2. The emission of this symmetric-top molecule is useful to probe temperatures and densities of hot molecular cores/corinos \citep{remijan2004}.  

\begin{figure*}
   \centering
    \includegraphics[width=18cm]{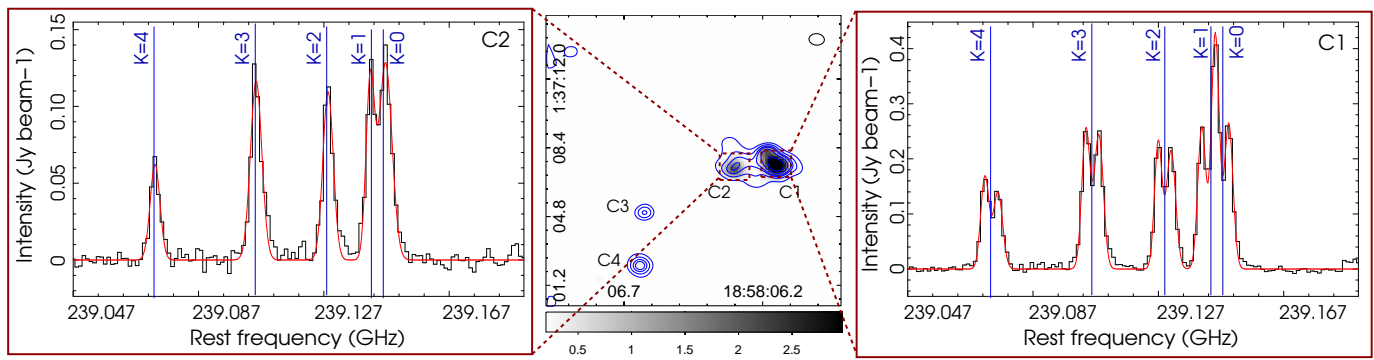}
    \caption{Center: integrated map of the CH$_{3}$CN J=13--12 emission (K=0,1 projections). Grayscale unit is Jy beam$^{-1}$~\ks. Blue contours represent the continuum emission at 334~GHz. Levels are at 15, 30, 50, 80, 140, and 200~mJy beam$^{-1}$. The beam is indicated at the top right corner. The spectra of the CH$_{3}$CN J=13--12 emission towards C2 and C1 are presented at the left and right panels, respectively. The red curves in both spectra indicate the Gaussian fittings.}
              \label{ch3cn}
\end{figure*}

Figure\,\ref{ch3cn}, center, shows the integrated emission map (K=0 and K=1 projections) of the CH$_3$CN J=13--12 transition. The blue contours represent the ALMA continuum emission at 334~GHz. The CH$_3$CN emission positionally coincides with the location of the cores C1 and C2. Figure \,\ref{ch3cn}, right and left, shows the average spectra towards the cores C1 and C2, respectively. The CH$_3$CN J=13--12 spectrum towards C1 exhibits conspicuous dips features in all K projections. A possible reason for such dip features might be the self-absorption. However, works like \citet{cesa14} that found an opacity of about 7 associated with the K=2 projection of the CH$_3$CN J=12--11 line without dip signatures, suggest that this possibility is unlikely. Therefore, we wonder if the spectrum behaviour could correspond to two velocity components associated with a possible rotating disk.

\begin{figure}
   \centering
    \includegraphics[width=9cm]{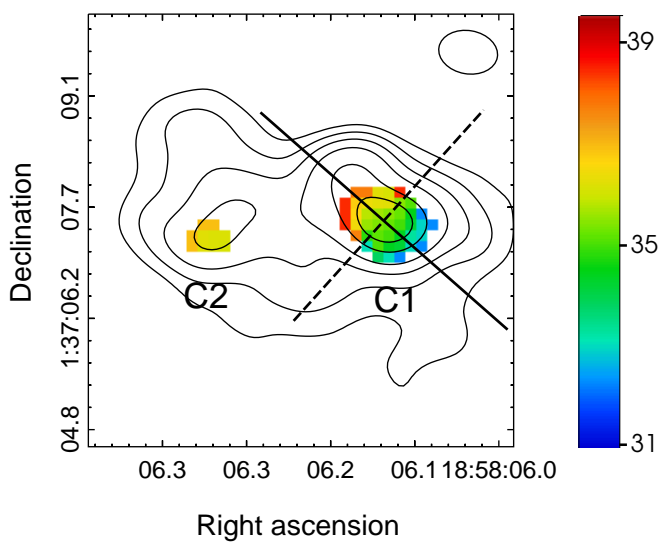}
    \caption{Moment 1 map of the CH$_{3}$CN J=13--12 emission (K=3 projection). Color scale unit is \ks. The black contours represent the continuum emission at 334~GHz. Levels are at 15, 30, 50, 80, 140, and 200~mJy beam$^{-1}$. The beam is indicated at the top right corner. The solid and dashed lines show the approximated directions of the disk and its rotation axis, respectively.}
              \label{ch3cn-mom1}
\end{figure}

Figure\,\ref{ch3cn-mom1} shows the moment 1 map of the CH$_{3}$CN J=13--12 emission (K=3 projection) towards the cores C1 and C2. It can be noticed a clear velocity gradient towards the core C1, which as was reported by several works \citep[e.g.,][]{louvet16}, would be indicating the presence of a rotating disk. The direction of the disk coincides with the direction in which the core C1 exhibits an elongated morphology. Moreover, in the case of rotating disks, it is expected to find symmetric spectral wings, which is particularly evident towards the projections K=2 and K=3 of the C1 spectrum (see Fig. \ref{ch3cn}-right). On the other hand, the behavior of the blended projections K=0 and K=1, whose most intense peak presents a larger intensity than the other projections, is better explained as the superposition of the K=0 blue component and the K=1 red component than as a consequence of self-absorption effects. 

The relevant tabulated numbers for the first five K projections of the CH$_3$CN J=13--12 transition are presented in Table\,\ref{CH3CN_LAB}. Columns 1 and 2 show the K projection and the rest frequency, respectively, obtained from NIST catalogue\footnote{https://physics.nist.gov/cgi-bin/micro/table5/start.pl}. Column 3 presents the upper energy level ($E_{u}/k$) extracted from the LAMDA database\footnote{https://home.strw.leidenuniv.nl/$~$moldata/} and column 4 shows the line strength of the projection multiplied by the dipole moment of the molecule ($S_{ul}\mu^{2}$).

\begin{figure}[h]
   \centering
   \includegraphics[width=9.5cm]{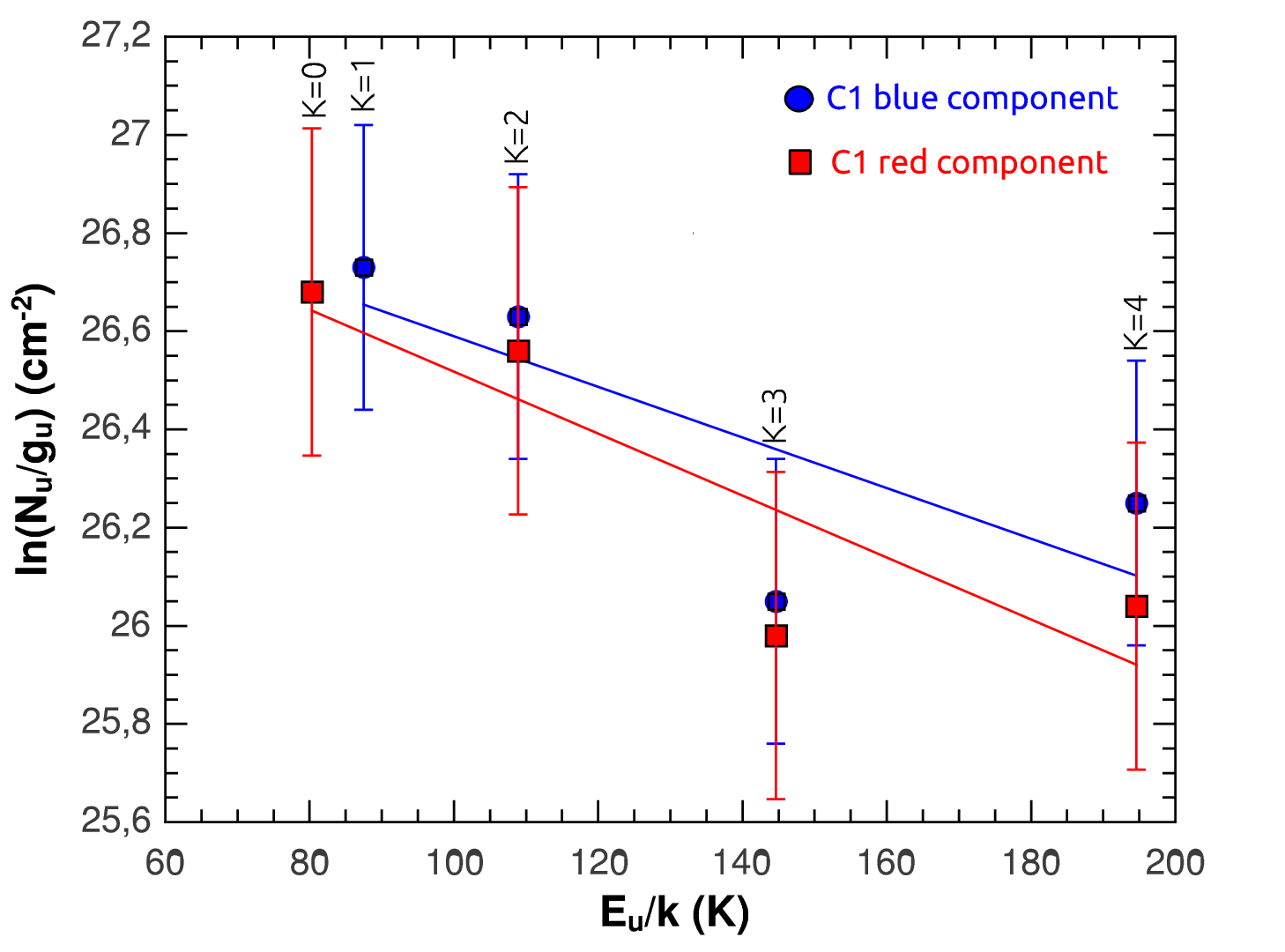}
    \includegraphics[width=9.5cm]{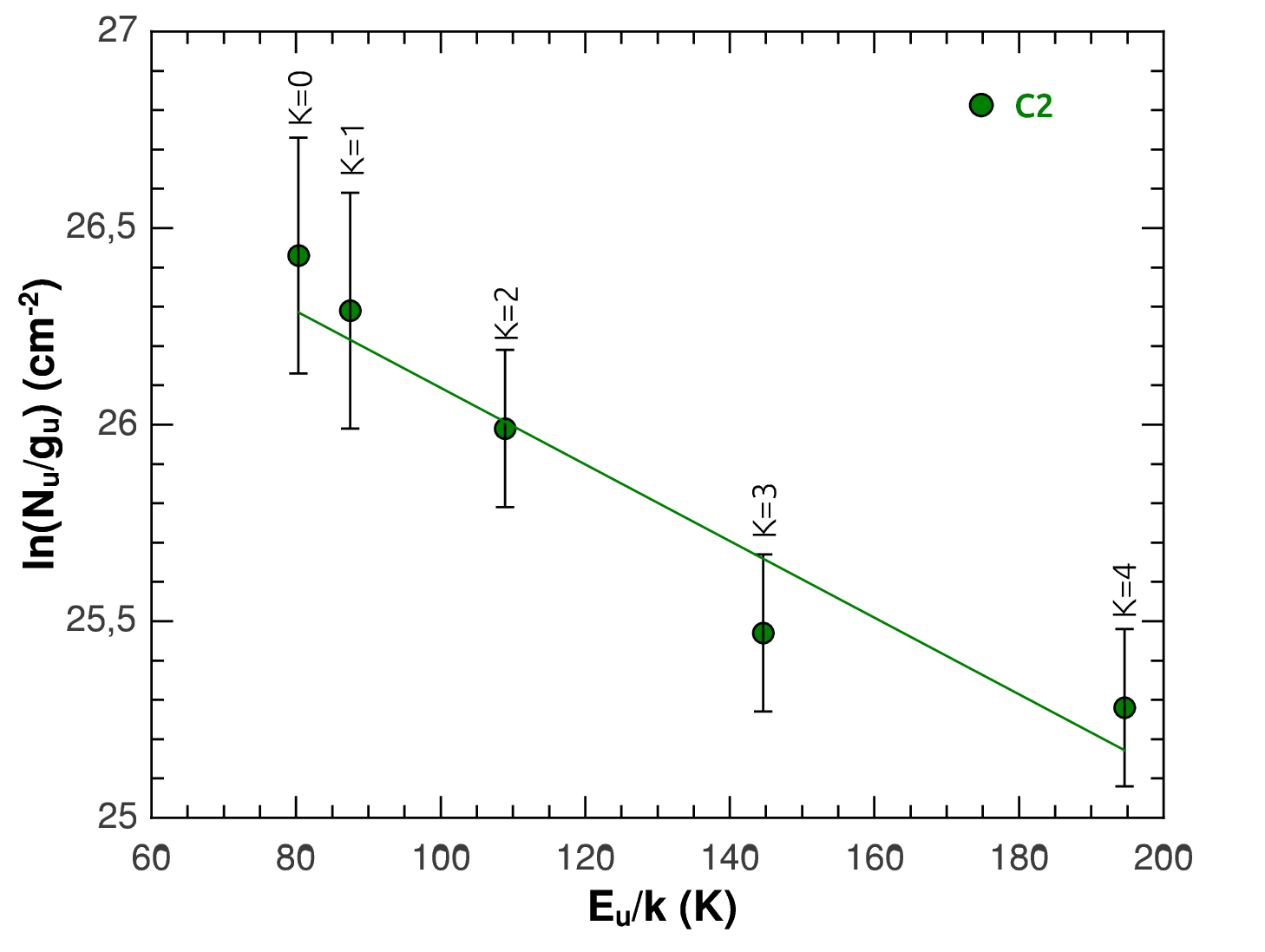}
       \caption{Rotational diagrams of the CH$_3$CN J=13$-$12 for the cores C1 (upper panel) and C2 (lower panel). The core C1 is separated in components blue and red. The blue component  of  the  K=0  projection  and  the  red  component  of  the K=1 projection are not fitted.}
              \label{Trot}
    \end{figure}

Table\,\ref{ch3cnKs} shows the main parameters derived from the  Gaussian fittings to the CH$_3$CN spectra of the cores C1 and C2. Columns 2, 3, and 4 show the peak intensity, the $\Delta$v, and the integrated intensity (W), respectively. The integrated intensities were used to construct the rotational diagram (RD) presented in Fig.\,\ref{Trot}. Thus, using the RD analysis \citep[][and references therein]{goldsmith99} and assuming LTE conditions, optically thin lines, and a beam filling factor equal to the unity, we can estimate the rotational temperature ($T_{rot}$) and the column density of the CH$_3$CN for the cores C1 and C2. This analysis is based on a derivation of the Boltzmann equation,

\begin{equation}
{\rm ln}\left(\frac{N_u}{g_u}\right)={\rm ln}\left(\frac{N_{tot}}{Q_{rot}}\right)-\frac{E_u}{kT_{rot}},   \label{RotDia} 
\end{equation}

\noindent where $N_u$ represents the molecular column density of the upper level of the transition, $g_u$  the total degeneracy of the upper level, $N_{tot}$ the total column density of the molecule, $Q_{rot}$ the rotational partition function, and $k$ the Boltzmann constant. 

Following \citet{miao95}, for interferometric observations, the left-hand side of Eq.\,\ref{RotDia} can also be estimated by,

\begin{equation}
{\rm ln}\left(\frac{N_u^{obs}}{g_u}\right)={\rm ln}\left(\frac{2.04 \times 10^{20}}{\theta_a \theta_b}\frac{W}{g_kg_l{\nu_0}^3S_{ul}{\mu_0}^2}\right),   
\label{RD} 
\end{equation}

\noindent where $N_u^{obs}$ (in cm$^{-2}$) is the observed column density of the molecule under the conditions mentioned above, $\theta_a$ and $\theta_b$ (in arcsec) are the major and minor axes of the clean beam, respectively, $W$ (in Jy beam$^{-1}$ \ks) is the integrated intensity of each K projection, $g_k$ is the K-ladder degeneracy, $g_l$ is the degeneracy due to the nuclear spin, $\nu_0$ (in GHz) is the rest frequency of the transition, $S_{ul}$ is the line strength of the transition, and $\mu_0$ (in Debye) is the permanent dipole moment of the molecule.  The free parameters, ($N_{tot}/Q_{rot}$) and $T_{rot}$ were determined by a linear fitting of Eq.\,\ref{RotDia} (see Fig.\,\ref{Trot}). We derive a $T_{rot}$ of about 190 and 170~K for the blue and red components of the core C1, respectively,  and of about 100~K for the core C2. Using the tabulated value for $Q_{rot}$ at the corresponding temperature, extracted from the CDMS database\footnote{https://cdms.astro.uni-koeln.de/cdms/portal/queryForm}, we obtain a CH$_3$CN column densities about 3.1 and 3.3~$\times 10^{15}$ cm$^{-2}$ for the blue and red components of the core C1, respectively, and $\sim1.2 \times 10^{15}$ cm$^{-2}$ for the core C2. 

If optical depths are high the measured line intensities will not reflect the column densities of the levels. Optical depth effects will be evident in the RD as deviations of the intensities from a straight line and a flattening of the slope, leading to anomalously large values for $T_{rot}$. The method, proposed by \citet{goldsmith99}, iteratively correct individual $N_u/g_u$ values by multiplying by the optical depth correction factor, $C_{\tau}=\tau/(1-e^{-\tau})$. However, we find that the $\tau$ corresponding to K=0 projection is about 0.002 and 0.003 for the cores C1 and C2, respectively, which leads to a correction factor less than 1 per cent. As expected, the others projections have correction factors even smaller.

\begin{table}
\caption{Tabulated parameters for the first five K projections of CH$_{3}$CN J=13--12.}
\centering
\begin{tabular}{c c c c}
\hline\hline
    Proj.     &   Rest frequency &  $E_u/k$ & $S_{ul}\mu^2$\\
   K       &   (GHz)   & (K)       &   (Debye$^2$) \\ 
   
\hline                 
    0     &  239.137    &  80.34 & 199.1\\
    1     &  239.133    &  87.49 & 197.9\\
    2     &  239.119    & 108.92 & 194.3\\
    3     &  239.096    & 143.63 & 188.5\\
    4     &  239.064    & 194.62 & 180.8\\ 
 \hline
 \hline
\end{tabular}
\label{CH3CN_LAB}
\end{table}

\begin{table}
\caption{Gaussian fittings parameters (see Fig.\,\ref{ch3cn}) for the K projections of the CH$_{3}$CN J=13--12 transition towards the cores C1 and C2. The core C1 is separated in components blue and red. The blue component of the K=0 projection and the red component of the K=1 projection for the core C1 are not fitted.}
\centering
\begin{tabular}{c c c c}
\hline\hline
    Proj.     &   Peak Int. & $\Delta$v & W \\
   K       &    (Jy beam$^{-1}$)  & \ks &(Jy beam$^{-1}$ \ks) \\
\hline
& & Core C1 (blue) &   \\    
\hline                 
    0     &  $-$ & $-$ & $-$  \\
    1     &  0.26 & 3.89 & 1.29  \\
    2     &  0.23 & 3.53 & 1.16  \\
    3     &  0.25 & 3.85 & 1.21  \\
    4     &  0.16 & 3.34 & 0.73 \\ 
\hline
  & & Core C1 (red) &   \\    
\hline                 
    0     &  0.26 & 4.11 & 1.24  \\
    1     &   $-$   &  $-$  &   $-$   \\
    2     &  0.22 & 4.29 & 1.07  \\
    3     &  0.24 & 4.19 & 1.17  \\
    4     &  0.14 & 3.91 & 0.60 \\ 
\hline
  & & Core C2  &   \\    
\hline                 
    0     &   0.13 & 6.1  &  0.96   \\
    1     &   0.12 & 4.0  &  0.83   \\
    2     &   0.11 & 4.9  &  0.61    \\
    3     &   0.12 & 5.4  &  0.70   \\
    4     &   0.06 & 4.6  &  0.28   \\
\hline
\hline
\end{tabular}
\label{ch3cnKs}
\end{table}

Finally, considering the integrated intensities at 334~GHz, assuming LTE conditions ($T_{rot} = T_{kin}$) with $T_{rot}$ of about 180 and 100~K for the cores C1 and C2, respectively, thermal coupling between dust and gas ($T_{kin}  = T_{dust}$), and following the same procedure presented in Sect.\,\ref{dust}, we re-estimate the masses of the cores C1 and C2 in about 1.4 and 0.9~\msol, respectively. These masses are about an order of magnitude smaller than those calculated using a temperature of 25~K (see Column 11 in Table \ref{contparams}). 

\subsection{Tracing the substructure of AGAL35 with other molecules} \label{molec_results}

\begin{figure}[h]
   \centering
   \includegraphics[width=8.5cm]{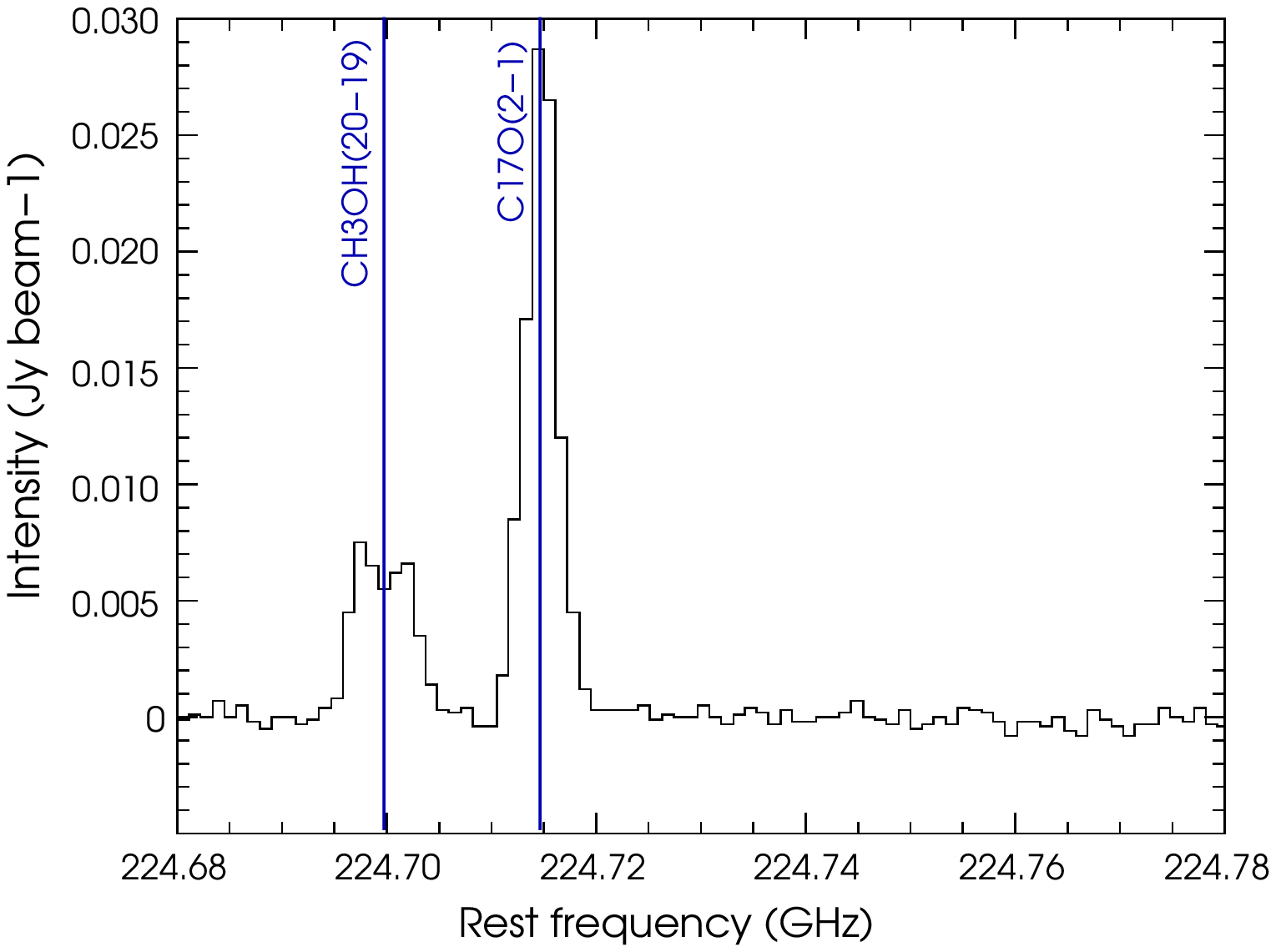}
   \includegraphics[width=8.5cm]{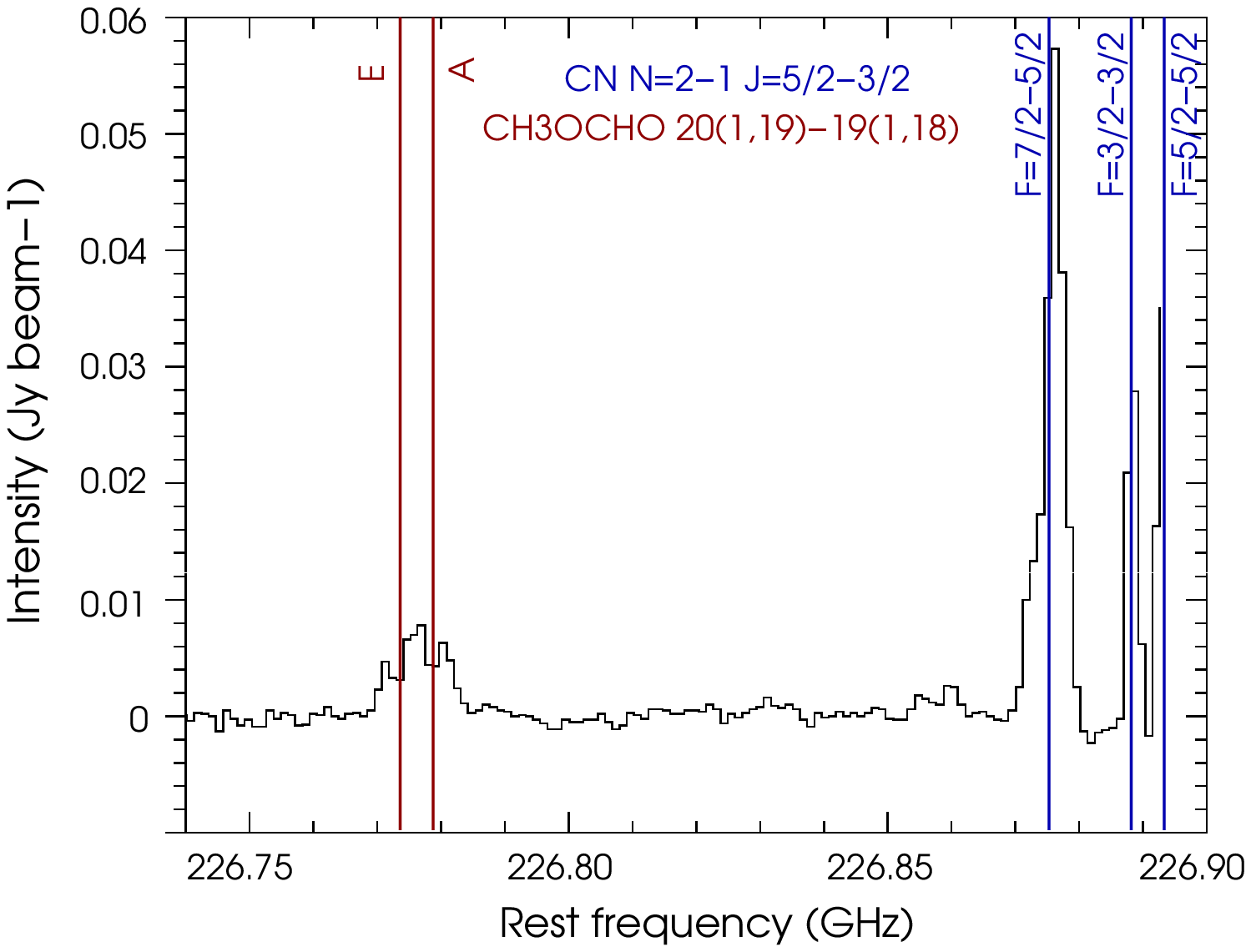}
       \caption{Spectra in the frequency ranges 224.68--224.78~GHz (up) and 226.74--226.90~GHz (bottom) obtained from a region of about 3\arcsec~in radius centered at the position of condensation MM1.}
              \label{C17OCN}
    \end{figure}

Figure\,\ref{C17OCN} shows spectra in the frequency ranges 224.68--224.78~GHz and 226.74--226.90~GHz obtained from a region of about 3\arcsec~in radius centered at the position of the condensation MM1. The C$^{17}$O (2--1), CH$_3$OH (20--19), CH$_3$OCHO (20--19) and CN (2--1) transitions are identified. 

\begin{figure}[h]
   \centering
   \includegraphics[width=9cm]{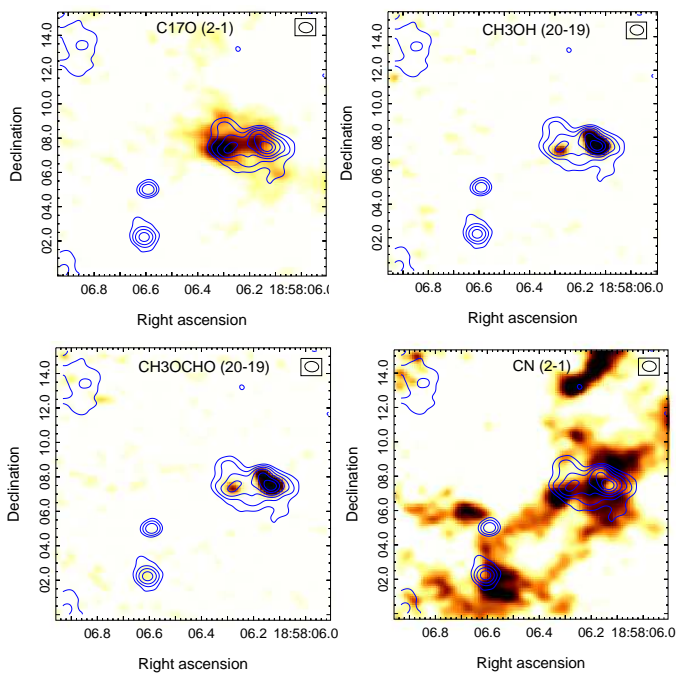}
    \caption{Integrated emission maps of C$^{17}$O (2--1), CH$_3$OH 20(-2,19)--19(-3,17)E,  CH$_3$OCHO 20(1,19)--19(1,18)  and  CN N=2--1, J=5/2--3/2, F=7/2--5/2 transitions. Color-scale goes from 0.05 to 0.4 Jy beam$^{-1}$~\ks in all panels. Blue contours represent the continuum emission at 334 GHz. Levels are at 15, 30, 50, 80, 140, and 200~mJy beam$^{-1}$. The beam is indicated at the top right corner of each panel.}
              \label{intmaps}
    \end{figure}

Figure\,\ref{intmaps} shows the integrated emission maps of C$^{17}$O (2--1), CH$_3$OH 20(-2,19)--19(-3,17)E,  CH$_3$OCHO 20(1,19)--19(1,18)(E-A)  and  CN N=2--1, J=5/2--3/2, F=7/2--5/2 transitions. The blue contours correspond to the continuum emission at 334~GHz. It can be noticed that the integrated emission of CH$_3$OH (20--19) and CH$_3$OCHO (20$-$19) spatially coincides with the position of the cores C1 and C2 at 334~GHz and no extended emission is detected above 5$\sigma$ rms noise level. 

On the other hand, C$^{17}$O (2--1) integrated map shows extended emission associated with the condensation MM1 (cores C1 and C2). The emission peak appears shifted about a beam size from the C2 continuum emission peak. A secondary emission peak is located towards the northeastern border of core C1. In addition, the CN (2$-$1) emission exhibits a clumpy and filamentary structure, which seems to connect all the cores. The CN molecular gas shows several condensations, which do not present positional coincidence with any of the cores. In fact, it can be noticed how the CN (2--1) emission distribution seems to surround the cores position, which is particularly evident towards C1.

\begin{figure}[h]
   \centering
   \includegraphics[width=9cm]{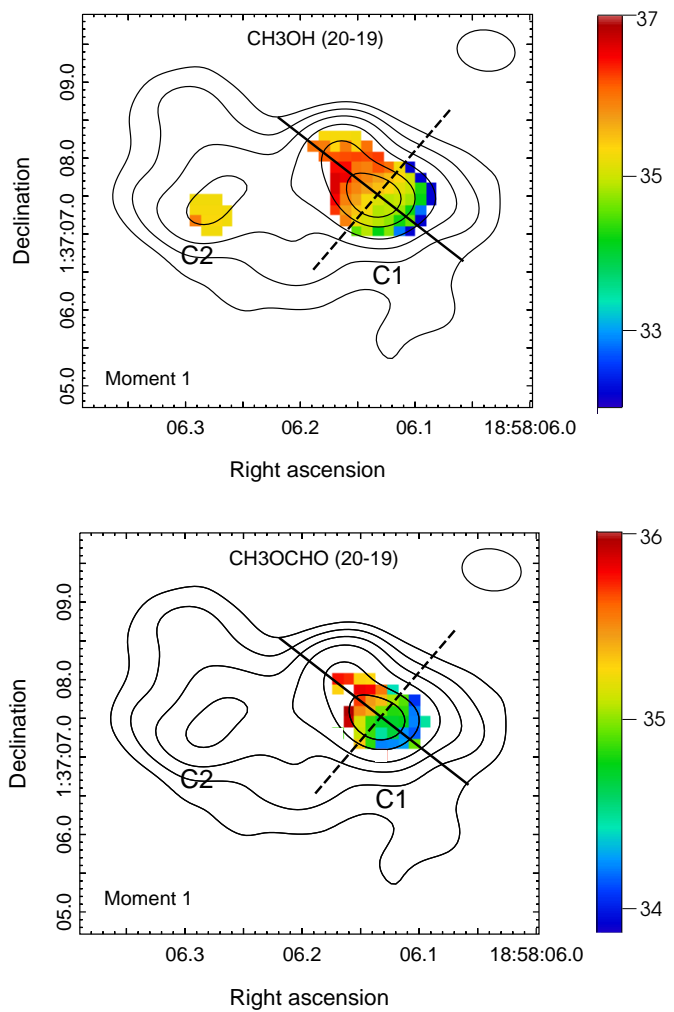}
    \caption{Moment 1 maps of CH$_3$OH 20(-2,19)--19(-3,17)E,  and CH$_3$OCHO 20(1,19)--19(1,18)   transitions. Color-scale unit is \ks. Black contours represent the continuum emission at 334 GHz. Levels are at 15, 30, 50, 80, 140, and 200~mJy beam$^{-1}$. The beam is indicated at the top right corner of each panel. The solid and dashed lines show the approximated directions of the disk and its rotation axis, respectively.}
              \label{molec-mom1}
    \end{figure}

Figure \ref{molec-mom1} shows the moment 1 maps for CH$_3$OH 20(-2,19)--19(-3,17)E and CH$_3$OCHO 20(1,19)--19(1,18) transitions. It can be appreciated clear velocity gradients towards the core C1 in both molecules, which is in agreement with what is observed with the CH$_3$CN (13-12) transition. This result reinforces the interpretation of a rotating disk towards C1.

\section{Discussion}
\label{discussion}

In this section we discuss the main results of our analysis of the substructure of AGAL35.

\subsection{Fragmentation of the massive clump AGAL35}

Massive clumps have a relatively low thermal Jeans mass of about 1~\msol~at typical densities (n $\sim 4.6~\times$ 10$^5$~cm$^{-3}$) and temperatures  (T $\sim$ 18~K), which predicts a high level of fragmentation. However, \citet{csengeri2017}, studying a homogeneous sample of infrared quiet massive clumps (in which it is included AGAL35), found limited fragmentation from clump to core scales with an average number of fragments of about 3 and with a mean mass of  about 63~\msol, which correspond to massive dense cores \citep{motte2007}. The authors suggested that early fragmentation of their massive clumps sample might not follow thermal processes, showing fragment masses largely exceeding the local Jeans mass. A possible explanation would be that the selected clumps with the highest peak surface density could correspond to a phase of compactness where the large level of fragmentation to form a cluster has not yet developed.  The authors suggested that a combination of turbulence, magnetic field, and radiative feedback could have increased the necessary mass for the fragmentation at the early stages. 

On the other hand, the authors explored the possibility of a global collapse occurring at cloud scale, in which the equilibrium does not be reached on small scales, which could lead to the observed limited fragmentation. Considering densities of about 10$^2$~cm$^{-3}$ at cloud scales, the initial thermal Jeans mass could reach 50~\msol, which is still not enough to explain the mass reservoir of the most massive cores of the sample. They also suggested that there might be mass replenishment beyond the clump scale that could fuel the formation of the lower mass population of stars, leading to an increase in the number of fragments with time and allowing a Jeans-like fragmentation to develop at more evolved stages. 



\citet{csengeri2017}, using the ALMA 7~m array at Band 7 with an angular resolution of about 4\arcsec, identified two cores MM1 and MM2, with masses of 36 and 8~\msol, respectively, towards AGAL35, which gives the total mass of fragments of 44~\msol. The higher angular resolution and sensitivity of the ALMA data presented in this work, allows us to identify four cores embedded within this massive clump. In particular, cores C1 and C2 correspond to the fragmentation of MM1 and cores C3 and C4 would correspond to the fragmentation of MM2. Assuming a temperature of 25~K, the same as in \citet{csengeri2017}, the total mass in fragments considering the four cores is about 25.7~\msol~($\sim$ 60\% of the MM1+MM2 mass) at 334~GHz and about 19.3~\msol~($\sim$ 45\% of the MM1+MM2 mass) at 239~GHz. The discrepancy of about a factor 2 between the total mass estimated in this work and those derived by \citet{csengeri2017}, might be due to the more extended array configuration of our data, which would not include part of the envelope emission in the mass estimation.

It is important to mention that from the CH$_3$CN J=13--12 emission, we find temperatures of about 180 and 100~K towards the cores C1 and C2, respectively. Moreover, the detection of the CH$_{3}$OH 20(-2,19)--19(-3,17)E line, with an energy $E_{u} = 514$ K, reinforces the temperature above 100~K as estimated from the CH$_3$CN emission. For instance, \citet{gins17} using several CH$_{3}$OH lines with frequencies and $E_{u}$ close to that presented here, obtained $T>100$ K in the W51 high-mass star-forming complex. Therefore, considering these temperatures, the summed mass of the cores C1 and C2 of about 2.3~\msol~(see Section \ref{CH3CN}) is about an order of magnitude smaller than the one reported by \citet{csengeri2017} towards the core MM1. Considering the parameters of the clump AGAL35 derived in previous works: a temperature of 25~K, a mass of 466~\msol, and a clump radius of 0.2~pc, we derive a Jeans mass, M${\rm _J}$, and a Jeans length, ${\rm \lambda_{\rm J}}$, of about 0.9~\msol~and 0.02~pc, respectively. Thus, the masses of the four cores that goes from 0.9 to 3.1~\msol~and their average separation that is about 0.04~pc, are in agreement with the expected Jeans mass and length for this clump.  

Our results suggest that the fragmentation of the massive clump AGAL35 gave rise to four low-intermediate mass cores, and that the major source of discrepancy in the masses estimation arises from the cores temperature. This study shows that in addition to the importance of carrying out high-resolution and sensitivity observations in order to have a complete detection of all fragments in a clump, it is very important to accurately determine the temperature of such cores for a correct estimation of their masses. 

On the other hand, given that our results suggest that the fragmentation of AGAL35 leads to low-intermediate mass cores, the only possibility for high mass star formation to take place in AGAL35 would be the competitive accretion scenario, in which the cores would increase their mass through material flowing through filaments. Observational studies of the kinematics of the molecular gas in star-forming regions \citep[e.g.,][]{sch19} have found the presence of velocity gradients in hub-filament systems on scales of 0.1 pc, possibly indicating mass transport at small scale. In section \ref{molecules}, we discuss briefly this possibility.

\subsection{The cores C3 and C4 as powering sources of MHOs}
\label{mho_disc}

The near-IR emission from  H$_2$ 1--0 S(1) line at 2.122~$\mu$m can be generated by either thermal emission in shock fronts or fluorescence excitation by non-ionizing UV photons \citep{martin2008, frank2014}, tracing the jets or the cavities carved out by the protostar jets and/or winds, respectively.  

\citet{froebrich2011}, using UWISH2 data, found several MHOs towards AGAL35, which shows that the region is an active site of star formation. The authors identified possible sources for some of these MHOs based on infrared color criteria (see Table \ref{mhos}). 

In this work, we suggest that the infrared-quite cores C3 and C4 are responsible for some MHOs in the region. Figure \ref{MHOs_CO} shows the perfect alignment among the molecular outflow OC4 and MHO2423 A and B, which suggests that the source responsible of these MHOs is the core C4. Additinally, the alignment among OC3 and MHO2426 A and B is striking, which points to the core C3 as the responsible source for these MHOs. Finally, MHO2424A is likely related to EGO35. 

\citet{samal2018} suggested that H$_2$ emission fades very quickly as the objects evolve from protostars to pre-main-sequence stars, which is in agreement with the relatively young dynamical age of about 10$^3$ yrs derived for the molecular outflows related to the cores C3 and C4. Moreover, to roughly check the evolutionary stage of the sources embedded in the cores C3 and C4, we looked for near-infrared point sources related to these cores in the UKIDSS catalog. The offset between the closest UKIDSS sources and the cores C3 and C4 is larger than 1\farcs5, which suggests that there is no spatial correlation between the near infrared sources and these submillimeter cores. The absence of near-infrared emission of the cores indicates that the star formation activity is going through early stages. On the other hand, the well collimated morphology exhibited by the molecular outflows labeled OC3 and OC4, reinforces the low-mass protostar scenario \citep{wu2004}.

\begin{table}
\centering
\caption{MHOs and their powering source candidates following \citet{froebrich2011}(see Fig. \ref{MHOs_CO}, left).}
\label{mhos}
\begin{tabular}{lc}
\hline\hline
    MHO         &  Powering source cand. \\
\hline\hline                 
2423 A & obj.\,D \\
2423 B & obj.\,D \\
2424 A & obj.\,A (Bright EGO) or E\\
2424 B & obj.\,E\\
2425   & obj.\,C\\
2426 A & obj.\,B (Faint EGO)\\
2426 B & obj.\,B (Faint EGO)\\
\hline
\end{tabular}
\end{table}

\subsection{Molecules in AGAL35}
\label{molecules}

In this section we discuss the information that the observed molecular lines give us regarding the cores embedded in the molecular clump AGAL35.

Figures\,\ref{ch3cn}\,and\,\ref{intmaps} show that the CH$_{3}$CN, CH$_{3}$OCHO, and CH$_{3}$OH are concentrated towards the cores C1 and C2, and they do not present extended emission. This fact shows that these molecular species are well tracers of hot molecular cores/corinos as found in previous works (e.g. \citealt{areal20,molet19, beltran2018}). It is known that these complex molecular species form in the dust grain surfaces, and when the temperature increases they thermally desorbe from the dust. Particularly, when the temperature of a molecular core reaches about 90 K, CH$_{3}$OH thermally desorbes from the grain mantles, and its gas-phase abundance is enhanced close to the protostars \citep{brown07}.

On the other hand, Fig.\,\ref{intmaps} shows that the CN (2--1) line traces clumpy extended emission in agreement with a recent work of \citet{paron21}, in which the cyano radical emission was analyzed with high-angular resolution ALMA data towards a sample of ten massive clumps. In particular, towards AGAL35, it is worth noting that the CN peaks do not spatially coincide with the continuum dust cores. Furthermore, it is clear how the CN (2--1) emission seems to surround the cores, which is particularly evident towards C1. This phenomenon was also observed in several other sources (see \citealt{paron21,zapata08,beuther04}). \citet{tassis12}, based on non-equilibrium chemistry dynamical models, found that the CN molecule is depleted at densities above 10$^5$~cm$^{-3}$. Thus, this molecule seems to not trace the densest cores and its emission would arise from the outer layers of very dense gaseous features. On the other hand, a complementary explanation is that the protostars are still very embedded in the cores and they do not generate enough UV photons to produce CN emission \citep{beuther04}.  In addition, it is worth mentioning that the missing flux coming from more extended spatial-scales that are filtered out by the interferometer could play an important role \citep{paron21}. 

Interestingly, the CN (2--1) emission towards AGAL35 shows a filamentary structure that seem to connect the four dust cores. These filaments might be tracing the remnant gas from a recent fragmentation that occurred in AGAL35 or they could be mapping the molecular material that is being accreted into the cores. Finally, we discard that the CN filaments can be related to molecular outflows because they do not coincide with neither outflow mapped in the $^{12}$CO emission nor the structures observed at the H$_{2}$ near-IR emission.

\section{Concluding remarks} 
\label{concl}

The massive clump AGAL G035.1330$-$00.7450 exhibits clear evidence of fragmentation, harbouring four low-intermediate mass infrared-quiet cores labeled from C1 to C4. Two of them, cores C3 and C4, show well collimated, young, and low mass molecular outflows. The main cores, C1 and C2, present CH$_3$CN emission, from which temperatures of about 180 and 100~K, respectively, are derived. Using these temperatures, we estimate masses of about 1.4 and 0.9~\msol, which allows us to conclude that these sources are hot corinos. Such masses are in agreement with the Jeans mass estimated for this massive clump and are about an order of magnitude smaller than the mass values derived in previous works, in which a clump scale temperature of about 25~K was assumed. 

This study confirms on the one hand, the relevance of high angular resolution and sensitivity observations to properly identify the degree of fragmentation of a clump, and on the other, it reveals the importance of an accurate determination of the cores temperatures to estimate their masses. Therefore, it would be interesting to extend this study to a considerable sample of clumps with similar characteristics to deeply analyse the phenomenon of limited fragmentation reported by several works.

Interestingly, the CH$_{3}$CN, CH$_{3}$OCHO and CH$_{3}$OH emission is concentrated towards C1 and C2. In particular, the moment 1 maps obtained from the emission of these three molecules, show clear velocity gradients consistent with the presence of a rotating disk towards the core C1.

Finally, given that all cores in AGAL35 present masses below 3~\msol, the only possibility that high-mass stars could form would be through the competitive accretion mechanism. In this regard, the filamentary structure exhibited by the CN emission might be tracing the gas falling towards the cores, which would be evidence of a competitive accretion scenario.

\begin{acknowledgements}

We thank the anonymous referee for her/his very useful comments. This work was partially supported by grant PICT 2015-1759 awarded by ANPCYT. M.O. and S.P. are members of the Carrera del Investigador Cient\'\i fico of CONICET, Argentina. A.M. and N.I. are doctoral and posdoctoral fellows of CONICET, Argentina. This work is based on the following ALMA data: ADS/JAO.ALMA $\#$ 2015.1.01312, and 2013.1.00960. ALMA is a partnership of ESO (representing its member states), NSF (USA) and NINS (Japan), together with NRC (Canada), MOST and ASIAA (Taiwan), and KASI (Republic of Korea), in cooperation with the Republic of Chile. The Joint ALMA Observatory is operated by ESO, AUI/NRAO and NAOJ.

\end{acknowledgements}

%
%

\bibliographystyle{aa}  
\bibliography{ref}
\IfFileExists{\jobname.bbl}{}
{\typeout{}
\typeout{****************************************************}
\typeout{****************************************************}
\typeout{** Please run "bibtex \jobname" to optain}
\typeout{** the bibliography and then re-run LaTeX}
\typeout{** twice to fix the references!}
\typeout{****************************************************}
\typeout{****************************************************}
\typeout{}
}
\label{lastpage}

\end{document}